\documentclass[]{aa}
\usepackage{lastpage}
\usepackage{graphicx}
\usepackage[varg]{txfonts}
\usepackage{natbib}
\bibpunct{(}{)}{;}{a}{}{,}
\usepackage{color}
\usepackage{mathrsfs}
\usepackage{amsbsy}
\usepackage{mathtools}
\usepackage{tabularx}
\usepackage{multirow}
\usepackage[mathlines]{lineno}
\allowdisplaybreaks

\usepackage[colorlinks=true,
            linkcolor=blue,     
            citecolor=blue,      
            urlcolor=blue,       
            filecolor=blue]{hyperref}
\begin{document} 

%=============================================
% 1st page
\title{Binary Evolution Pathways to Blue Large-Amplitude Pulsators: Insights from HD 133729}

\author{Zhengyang Zhang\inst{1,2,3}\fnmsep
\and Chengyuan Wu\inst{1,2,3}
\and Xianfei Zhang\inst{4,5}
\and Zhanwen Han\inst{1,2,3}
\and Bo Wang\inst{1,2,3}
}
\institute{Yunnan Observatories, Chinese Academy of Sciences, Kunming 650011, People’s Republic of China
\email{wuchengyuan@ynao.ac.cn, wangbo@ynao.ac.cn}
\and
University of the Chinese Academy of Sciences, 19A Yuquan Road, Shijingshan District, Beijing 100049, People’s Republic of China
\and
International Centre of Supernovae, Yunnan Key Laboratory, Kunming, 650216, People’s Republic of China
\and
Institute for Frontier in Astronomy and Astrophysics, Beijing Normal University, Beijing 102206, People’s Republic of China
\and
Department of Astronomy, Beijing Normal University, Beijing 100875, People’s Republic of China
}

\date{Received; Accepted}

\abstract{
Blue Large-Amplitude Pulsators (BLAPs) represent a recently identified class of pulsating stars distinguished by their short pulsation periods ($2$ – $60$ minutes) and asymmetric light curves. This study investigated the evolutionary channel of HD 133729 which is the first confirmed BLAP in a binary system. Using the binary evolution code MESA, we explored various mass ratios and initial orbital periods. Our simulations suggest that a system with a mass ratio  $q = 0.30$ undergoing non-conservative mass transfer ($\beta=0.15$) can reproduce the observed characteristics through the pre-white dwarf Roche lobe overflow channel. Meanwhile, we predict that there are significant helium and nitrogen enhancements on the surface of the main sequence (MS) star. The system will eventually undergo the common envelope phase, leading to a stellar merger. HD 133729 is a unique case as a benchmark, providing crucial insights into the formation mechanism and evolutionary fate of BLAPs with MS companions. This work constrains the elemental abundances of the MS star and has helped our understanding of non-conservative mass transfer in binary evolution.
}

\keywords{evolution -- binaries: close -- stars: oscillations -- stars:}

\titlerunning{Binary Evolution of HD 133729}
\authorrunning{Zhengyang Zhang et al.}

\maketitle

%=============================================
% Introduction section with proper math formatting
% Introduction section with corrected math formatting
\section{Introduction} \label{intro}

Blue Large-Amplitude Pulsators (BLAPs) are a class of pulsating stars that was first observed by the Optical Gravitational Lensing Experiment (OGLE) survey in 2013 \citep[e.g.][]{2013AcA....63..379P}. BLAPs are characterized by short pulsation periods (22 to 40 minutes) and asymmetric light curves. These light curves show a rapid rise followed by a slower decline in brightness \citep[e.g.][]{2017NatAs...1E.166P}. BLAPs occupy a region on the Hertzsprung-Russell (HR) diagram between hot massive main sequence (MS) stars and hot subdwarfs. Their luminosities (${\rm log}(L/\rm L_{\sun}) = 2.2 - 2.6$) are approximately an order of magnitude higher than those of hot subdwarfs, whereas their surface gravities (${\rm log}(g/\rm cm \, s^{-2}) = 4.5-4.8$) are significantly lower. Besides typical BLAPs, there is also a class of high-gravity BLAPs, in which their spectroscopic properties and pulsation periods are more similar to those of sdB stars pulsating in the p-mode, with $T_{\rm eff} \simeq 32,000$K and pulsation periods between 200 and 475 seconds \citep[e.g.][]{2019ApJ...878L..35K}. \citet{2022MNRAS.511.4971M} recently combined photometric data from Gaia and time-series data from the Zwicky Transient Facility (ZTF), and reported the discovery of 22 BLAP candidates, including six high-gravity pulsators.

The pulsation mechanism of BLAPs is attributed to the $\kappa$ mechanism, driven by the iron opacity bump at a specific temperature. The phenomenon of radiative levitation arises from the outward force exerted by stellar radiation on ions within a star. This phenomenon was originally proposed to explain the pulsations observed in hot subdwarf stars \citep[e.g.][]{1996ApJ...471L.103C}. \citet{2018MNRAS.481.3810B} highlighted the significant role of radiative levitation, which leads to the accumulation of iron and nickel around the temperature of $\simeq 2 \times 10^{5}$ K. This accumulation forms a substantial opacity bump that drives the pulsations observed in BLAPs.

\begin{figure}[ht!]
\begin{center}
\resizebox{\hsize}{!}{\includegraphics{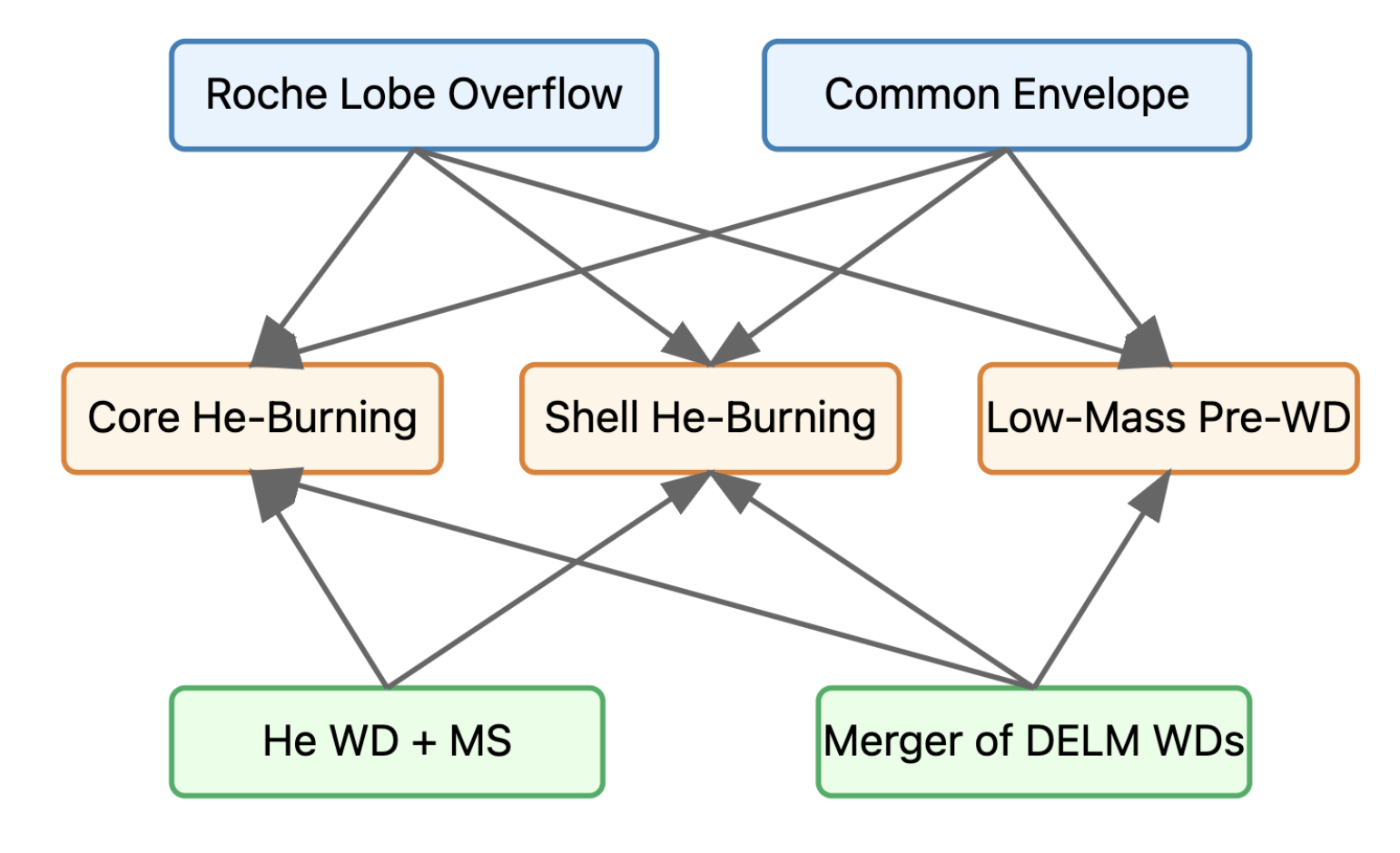}}
\end{center}
\caption{Schematic diagram for the origin of BLAPs. The starting point of the arrow indicates the stellar evolutionary pathway, while the end represents the star's evolutionary state.}
\label{fig1}
\end{figure}

Several models have been proposed to explain the origin of BLAPs (also see Figure \ref{fig1}):
\begin{enumerate}
\item Low-Mass pre-WD Model: This model suggests that BLAPs are low-mass stars (typically 0.2 - 0.4$\rm \, M_{\sun}$) with helium cores. Their high effective temperatures and luminosities are sustained by residual hydrogen shell burning, and they are formed via significant mass loss from a red giant star in binary systems, either through common envelope (CE)  evolution or Roche lobe overflow (RLOF) \citep[e.g.][]{2018MNRAS.477L..30R, 2018MNRAS.481.3810B, 2018arXiv180907451C, 2021MNRAS.507..621B}. \\

\item Helium-Burning Models: a) Core Helium-Burning Star. \cite{2018MNRAS.478.3871W} proposed that core helium-burning stars could match the observed properties of BLAPs. These stars evolve from initial masses greater than $4.0\rm \, M_{\sun}$ to final masses between 0.5 and 1.2$\rm \, M_{\sun}$. The common envelope wind (CEW) model, as proposed by \citet{2020ApJ...903..100M}, suggests that surviving companions of SNe Ia could evolve into helium-burning stars with thin hydrogen envelopes. b) Shell Helium-Burning Star. BLAPs may be shell helium-burning subdwarf B-type (SHeB sdB) stars, formed in long-period binary systems via stable RLOF \citep[e.g.][]{2022A&A...668A.112X}. \\

\item Stellar mergers can lead to the formation of BLAPs. For example, the merger of a He WD with a low-mass MS star can produce hot subdwarfs that pass through BLAP states between helium shell ignition and core burning \citep[e.g.,][]{2023ApJ...959...24Z}. The merger of double extremely low-mass (DELM) WDs is another potential formation channel for BLAPs. \citet{2024A&A...691A.103K} propose that DELM WD mergers with total system masses between 0.32 and 0.7$\rm \, M_{\sun}$ produce BLAPs. After coalescence, these systems quickly evolve into the BLAP phase. They remain in this state for 20,000 to 70,000 years before evolving into hot subdwarfs. Eventually, these systems evolve into either He WDs or hybrid He/CO WDs. This formation scenario explains the recent discovery of magnetic BLAPs, as strong magnetic fields are generated during the merger process.

\end{enumerate}

Although more than 80 BLAPs have been discovered, very few have been confirmed as binary systems \citep[e.g.][]{2024arXiv240416089P}. TMTS-BLAP-1 is potentially a wide binary system with an orbital period of 1576 days, but has not yet been confirmed \citep[e.g.][]{2023NatAs...7..223L}. Recent photometric analysis has revealed that HD 133729 is a binary system. It consists of a late B-type MS star and a BLAP companion. This makes HD 133729 the first, and currently the only confirmed BLAP in a binary system. Analysis of TESS data from both sectors revealed a dominant frequency at $f_{\rm P} = 44.486 \, \rm d^{-1}$, corresponding to a pulsation period of 32.37 minutes. Harmonics were also observed, indicating a non-sinusoidal light curve, a characteristic feature of BLAPs. The light-travel-time effect manifests as orbital sidelobes in the frequency spectra. These sidelobes appear at a frequency separation of $f_{\rm orb}$, the orbital frequency. For HD 133729, the sidelobe separation of approximately $0.043$ d$^{-1}$ implies an orbital period of $P_{\rm orb} = f_{\rm orb}^{-1} \approx 23$ days \citep[e.g.][]{2022A&A...663A..62P}. Additional observational data are provided in Table 1.

\begin{table}
\caption{Observational Data of HD 133729.}
\label{table1}
\centering
\begin{tabular}{c c c}         
\hline\hline
 & BLAP & B-type MS \\
\hline
$L/\rm L_{\sun}$ & $2.15 \pm 0.15$ & $1.87 \pm 0.12$ \\
$T_{\rm eff}$/K & $11,500 \pm 1000$ & $29,000 \pm 1800$ \\
${\rm log}(g/\rm cm \, s^{-2})$ & 4.5 & 4.0 \\
$M/\rm M_{\sun}$ & - & $2.85 \pm 0.25$\\
\hline                                          
\end{tabular}
\end{table}

Several unresolved questions remain regarding the formation of the BLAP in the HD 133729 system. Key questions include: Which formation channel led to its origin — pre-WD or helium-burning star? What is the evolution of the binary system after the BLAP enters the pulsational instability strip? How do the surface abundances of the B-type MS star compare to those of single MS stars? Could any abundance anomalies be used as diagnostic tools in spectroscopic analyses? Finally, was the mass transfer in HD 133729 conservative or non-conservative? HD 133729 offers a unique opportunity to investigate the formation and evolution of BLAPs. This study focuses on the evolutionary pathways of BLAPs with MS companions. HD 133729 serves as a prototype for this investigation.

This paper is structured as follows. In section \ref{method}, we show the methods and parameters used for our stellar evolution and pulsation calculations. Section \ref{result} presents the evolutionary outcomes, categorized by mass ratio. A discussion of these results is provided in Section \ref{diss}. Finally, Section \ref{sum} summarizes the key findings of this study.

\section{Methods and numerical input} \label{method}
To investigate the origin of HD 133729, we use MESA (version r23.05.1) to simulate the evolution of binary systems \citep[e.g.][]{2011ApJS..192....3P, 2013ApJS..208....4P, 2015ApJS..220...15P, 2018ApJS..234...34P, 2019ApJS..243...10P}. The "binary" module in MESA evolves binary systems by creating two single stars using the "star" module. This module can evolve a primary star with a point-mass companion or evolve both stars simultaneously. The latter approach was adopted, where both stars are co-evolved \citep[e.g.][]{2015ApJS..220...15P}.

First, we create a series of pre-MS stellar models using the \texttt{create\_pre\_main\_sequence\_model} in the "star" module, as shown in Table 2. We used a metallicity of $Z=0.02$ for all models. For opacity, we selected the GS98 table \citep[e.g.][]{1998SSRv...85..161G}. For convection, we applied the standard mixing length theory (MLT) with $\rm \alpha_{MLT}=2$ \citep{2024ApJ...975..186Z}. We did not include the effects of convective overshooting, thermohaline mixing, or semiconvection.

We then added the primary and secondary stars to the "binary" module. The mass transfer process follows the Ritter prescription \citep[e.g.][]{1988A&A...202...93R}. This prescription predicts that matter flows through the inner Lagrangian point L1 as an isothermal, subsonic gas stream. As the stream approaches L1, it reaches the speed of sound.

We examined both conservative and non-conservative evolution. In the conservative case, the system transfers mass between stars without additional mass loss from rapid stellar winds. In the non-conservative evolution case, mass is lost from the system alongside the interstellar mass transfer. In systems with low mass transfer efficiency, angular momentum loss is described by the model of \citet{1997A&A...327..620S}, in which a fixed fraction of the transferred mass is expelled either as fast, isotropic stellar winds from each star or as circumbinary material forming a ring at a specified radius. The accreted fraction of mass is defined as $\epsilon_w\equiv1-\alpha-\beta$. Here, $\alpha$ represents the fraction of mass lost as stellar winds and $\beta$ is the fraction isotropically re-emitted. We tested various $\beta$ values to analyze their impact on mass evolution and to match observed properties.

The pulsation period was calculated using the following formula:
\begin{equation}
M=\frac{10^{\langle\log(g)\rangle}R^2}{\rm G}=\frac{10^{3\langle\log(g)\rangle}f^4}{{\rm G}\omega^4}.
\end{equation}
This equation relates the mass of the star to its surface gravity, radius, and pulsation frequency. The parameter $f=\omega/\omega_{\rm dyn}$, where $\omega_{\rm dyn}^2=GM/R^3$ is the dynamical frequency of the star. For low-mass helium-core pre-WD models, $f \approx 3.6$. G is the gravitational constant \citep[e.g.][]{2019ApJ...878L..35K}.

To compare the accuracy of formulas for calculating pulsation periods, we used GYRE to compute the $l=0$ radial pulsations of stellar evolutionary models. In the Eckart-Scuflaire-Osaki-Takata scheme, the meaning of $n_{\rm pg}$ is as follows:
\begin{equation}
n_{\mathrm{pg}} = 
\begin{cases}
n_{\mathrm{p}}-n_{\mathrm{g}} & \text{for a } g_{n_{p}-n_{g}} \text{ mode}, \\
n_{\mathrm{p}}-n_{\mathrm{g}}+1 & \text{for a } p_{n_{p}-n_{g}+1} \text{ mode}.
\end{cases}
\end{equation}

The pulsations of BLAPs are generally radial fundamental modes. Therefore, we only focus on the case where $n_{\rm pg}=1$ \citep[e.g.][]{2006PASJ...58..893T}.

\citet{2021MNRAS.507..621B} found no strong correlation between the initial and final masses of BLAPs. This finding indicates that other binary parameters, such as the mass ratio and the initial orbital period, play a more significant role in determining the final mass. Consequently, we computed evolutionary models over various mass ratios and orbital periods.

\begin{table*}
\label{table2}
\caption{Initial parameters of binary evolution. The mass ratio of the secondary star to the primary star is $q = M_{\rm 2,i}/M_{\rm 1,i}$. $P_{\mathrm{minutes}}$ is the minimum initial period of the binary system. $P_{\mathrm{max}}$ is the maximum initial period of the binary system. $\Delta P$ is the step size of the orbital period. $\beta$ is the fraction of isotropically re-emitted.}
\centering
\begin{tabular}{c c c c c c c}
\hline\hline
$q$ & $M_{\rm 1,i}(\rm M_{\sun})$ & $M_{\rm 2,i}(\rm M_{\sun})$ & $P_{\mathrm{minutes}}(\rm day)$ & $P_{\mathrm{max}}(\rm day)$ & $\Delta P(\rm day)$ & $\beta$ \\
\hline
     & 2.32 & 0.58 & 1.0 & 4.0 & 0.3 &  0 \\
     & 2.40 & 0.60 & 1.0 & 4.0 & 0.3 &  0 \\
0.25 & 2.48 & 0.62 & 1.0 & 4.0 & 0.3 &  0 \\
     & 2.56 & 0.64 & 1.0 & 4.0 & 0.3 &  0 \\
     & 2.64 & 0.66 & 1.0 & 4.0 & 0.3 &  0 \\
     & 2.72 & 0.68 & 1.0 & 4.0 & 0.3 &  0 \\
\hline
     & 2.23 & 0.67 & 1.0 & 4.0 & 0.3 &  0 \\
     & 2.31 & 0.69 & 1.0 & 4.0 & 0.3 &  0 \\
0.30 & 2.38 & 0.72 & 1.0 & 4.0 & 0.3 &  0 \\
     & 2.46 & 0.74 & 1.0 & 4.0 & 0.3 &  0 \\
     & 2.54 & 0.76 & 1.0 & 4.0 & 0.3 &  0 \\
     & 2.62 & 0.78 & 1.0 & 4.0 & 0.3 &  0 \\
\hline
     & 2.62 & 0.78 & 2.9 & 3.5 & 0.2 &  0.05 \\
     & 2.62 & 0.78 & 2.9 & 3.5 & 0.2 &  0.10 \\
0.30 & 2.62 & 0.78 & 2.9 & 3.5 & 0.2 &  0.15 \\
     & 2.62 & 0.78 & 2.9 & 3.5 & 0.2 &  0.20 \\
\hline     
     & 2.14 & 0.75 & 1.0 & 4.0 & 0.3 &  0 \\
     & 2.22 & 0.78 & 1.0 & 4.0 & 0.3 &  0 \\
0.35 & 2.30 & 0.80 & 1.0 & 4.0 & 0.3 &  0 \\
     & 2.37 & 0.83 & 1.0 & 4.0 & 0.3 &  0 \\
     & 2.44 & 0.86 & 1.0 & 4.0 & 0.3 &  0 \\
     & 2.52 & 0.88 & 1.0 & 4.0 & 0.3 &  0 \\
\hline
     & 2.00 & 0.90 & 1.0 & 4.0 & 0.3 &  0 \\
     & 2.07 & 0.93 & 1.0 & 4.0 & 0.3 &  0 \\
0.45 & 2.14 & 0.96 & 1.0 & 4.0 & 0.3 &  0 \\
     & 2.21 & 0.99 & 1.0 & 4.0 & 0.3 &  0 \\
     & 2.28 & 1.02 & 1.0 & 4.0 & 0.3 &  0 \\
     & 2.35 & 1.05 & 1.0 & 4.0 & 0.3 &  0 \\
\hline 
\end{tabular}
\end{table*}

\section{Results}\label{result}
The formation and evolution of the BLAP progenitor in HD 133729 can be explained by two binary formation channels: the pre-WD RLOF/CE channel, and the helium-burning RLOF/CE channel. Both channels involve binary interaction. In HD 133729, the B-type MS star has a mass of $(2.85 \pm 0.25) \, \rm M_{\sun}$, which indicates that the original BLAP progenitor must have been the primary.

If the progenitor evolved through the helium-burning channel, it would have undergone significant mass loss. Since a ZAMS star requires at least $4 \, \rm M_{\sun}$ to eventually become a BLAP, roughly $3.5 \, \rm M_{\sun}$ would need to be removed to form a helium-burning star. Under the assumption of conservative mass transfer, the secondary star should have accreted enough mass to exceed $3.5 \, \rm M_{\sun}$. This contradicts the observed mass of the B-type MS star, rendering the helium-burning channel unlikely.

Our calculations further reveal that, for the pre-WD channel, systems with initial orbital periods longer than 100 days exhibit a rapid increase in the mass transfer rate—exceeding $\rm 10^{-3} \, M_{\sun} \, yr^{-1}$—once the donor fills its Roche lobe. This rapid mass transfer triggers CE. In these scenarios, the minimum required common envelope efficiency ($\rm \alpha_{CE}$) for successful envelope ejection exceeds 10, implying that these systems will likely merge. Therefore, our study focuses exclusively on the RLOF pre-WD channel.

\subsection{Cases for $q = 0.25$}
For $q = 0.25$ and an initial orbital period of 1.0 day, the primary star begins mass transfer while still on the MS. Its core hydrogen abundance is between 0.17 and 0.21, depending on the masses of the binary components. During this phase, the mass transfer timescale ($\tau_{\rm MT}\approx 10^{6.5}\,\rm yr$) gradually decreases and eventually becomes shorter than the thermal timescale ($\tau_{\rm KH}\approx 10^{6}\,\rm yr$). Nonetheless, mass accretion can still be maintained.

\begin{figure}[ht!]
\begin{center}
\resizebox{\hsize}{!}{\includegraphics{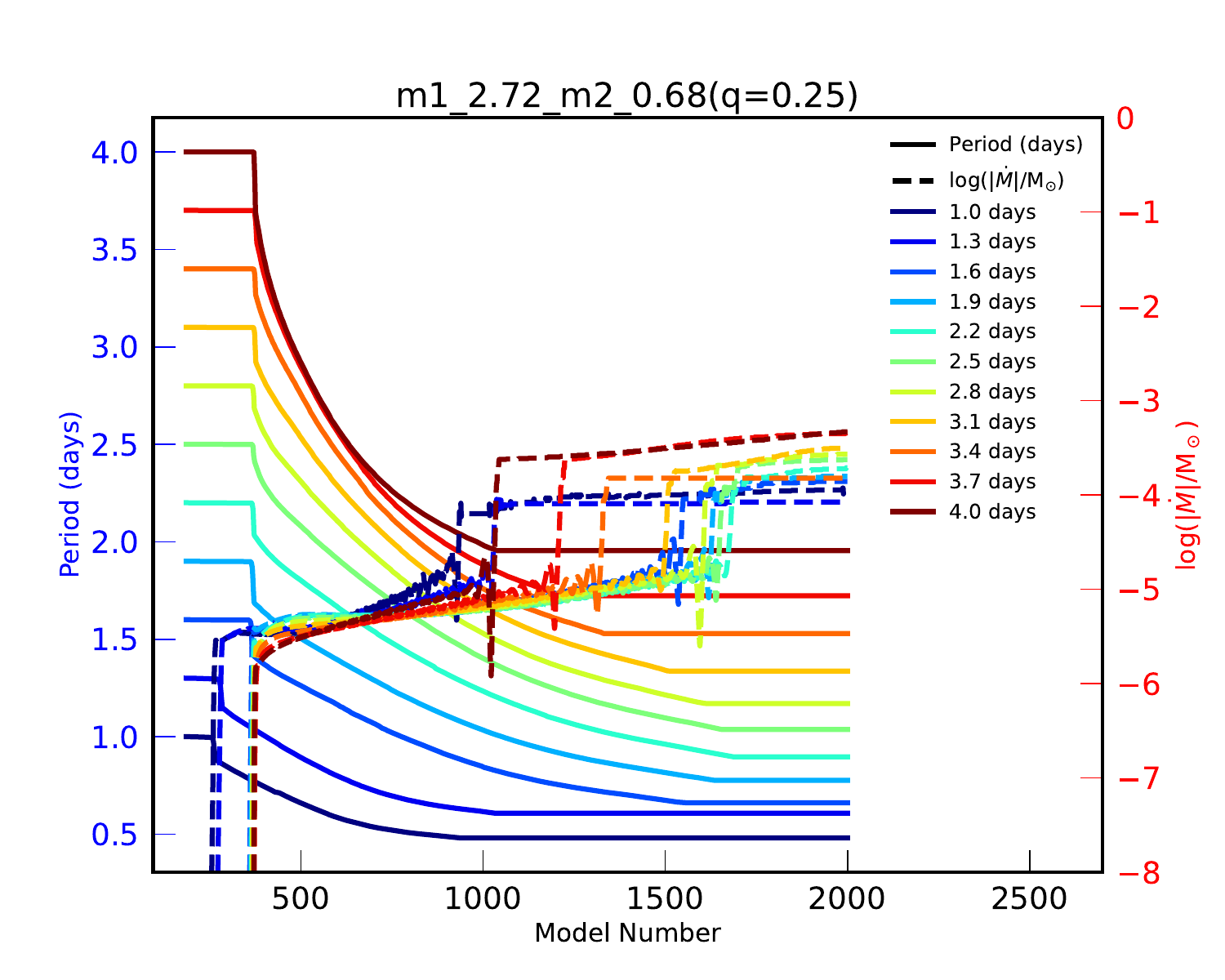}}   
\end{center}
\caption{Evolution tracks of a binary system consisting of a $M_{1} = 2.72 \rm \,  M_{\sun}$ primary and a $M_{2} = 0.68 \rm \, M_{\sun}$ companion. The x-axis shows the model number. Solid lines represent the orbital period (left y-axis) and dashed lines show the logarithm of the mass transfer rate (right y-axis). Different colors correspond to initial orbital periods ranging from 1.0 to 4.0 days.}
\label{fig2}
\end{figure}

When the orbital period decreases to approximately 0.5 days, the mass transfer timescale drops suddenly to $\tau_{\rm MT}\approx 10^{4.2} \,\rm yr$. This further increases the discrepancy between $\tau_{\rm MT}$ and $\tau_{\rm KH}$. As illustrated in Figure \ref{fig2}—which shows cases with the highest binary masses—the mass transfer rate becomes extremely high, around $10^{-4} \,\rm M_{\odot} \,yr^{-1}$. Such a rapid rate may trigger a CE phase.

A longer orbital period delays the onset of mass transfer. This delay leads to a lower hydrogen abundance in the primary star. However, as in the case with an orbital period of 1.0 days, the accretion rate increases dramatically once the orbital period shrinks to its minimum. This sudden increase suggests that the system is transitioning to a CE phase. Figure \ref{fig2} shows that even when the orbital period is increased to 4 days, the system behaves similarly. In this scenario, the accretion rate abruptly rises to nearly $\rm 10^{-3} M_{\odot} \, yr^{-1}$. This result suggests that systems with a mass ratio of 0.25 cannot undergo stable mass transfer, thus excluding this scenario.

\subsection{Cases for $q = 0.35$}

\begin{figure}[ht!]
\begin{center}
\resizebox{\hsize}{!}{\includegraphics{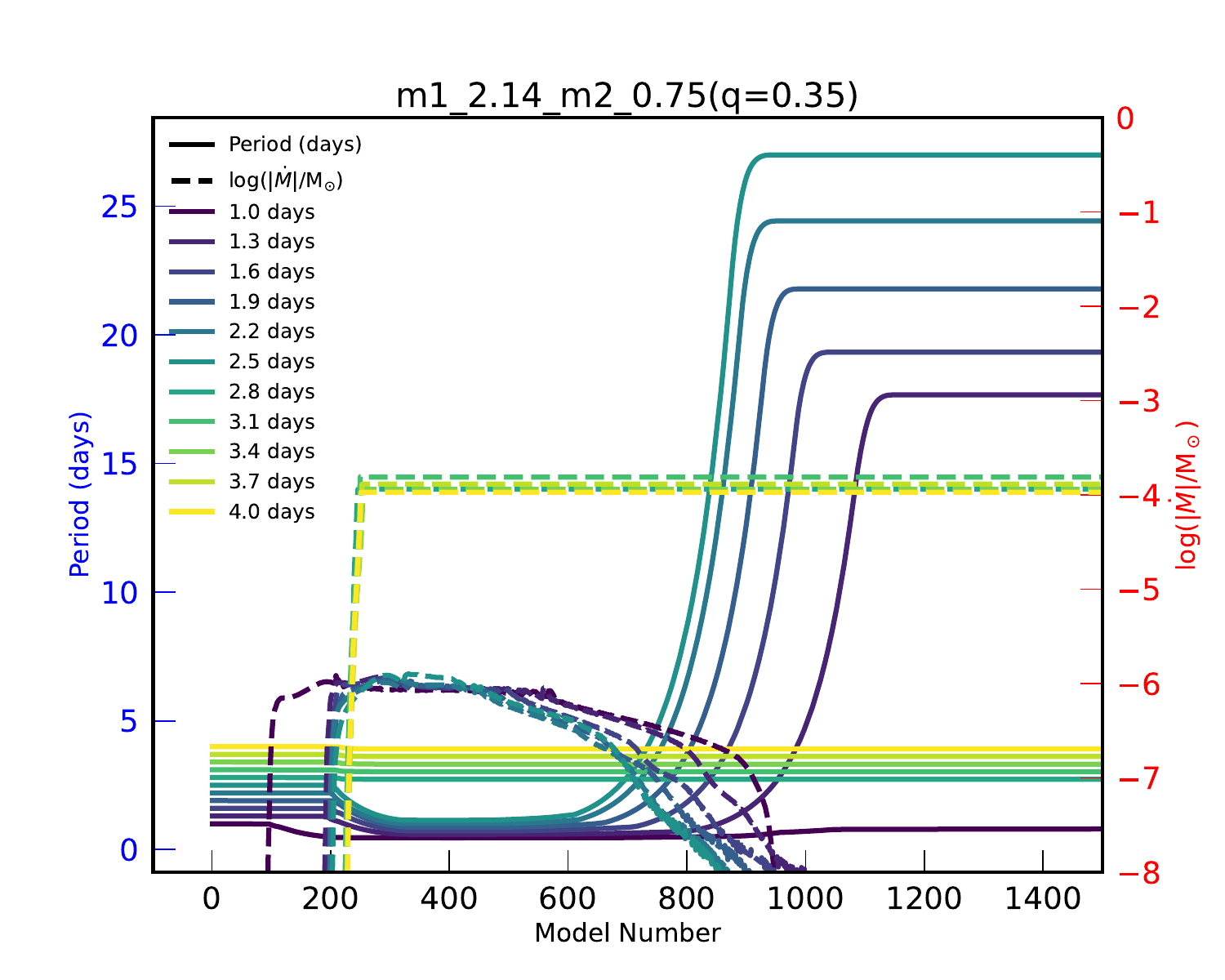}}    
\end{center}
\caption{Evolution tracks of a binary system consisting of a $M_{1} = 2.14 \rm \, M_{\sun}$ primary and a $M_{2} = 0.75 \rm \, M_{\sun}$ companion. The x-axis shows the model number. Solid lines represent the orbital period (left y-axis) and dashed lines show the logarithm of the mass transfer rate (right y-axis). Different colors correspond to initial orbital periods ranging from 1.0 to 4.0 days.}
\label{fig3}
\end{figure}

\begin{figure}[ht!]
\begin{center}
\resizebox{0.9\hsize}{!}{\includegraphics{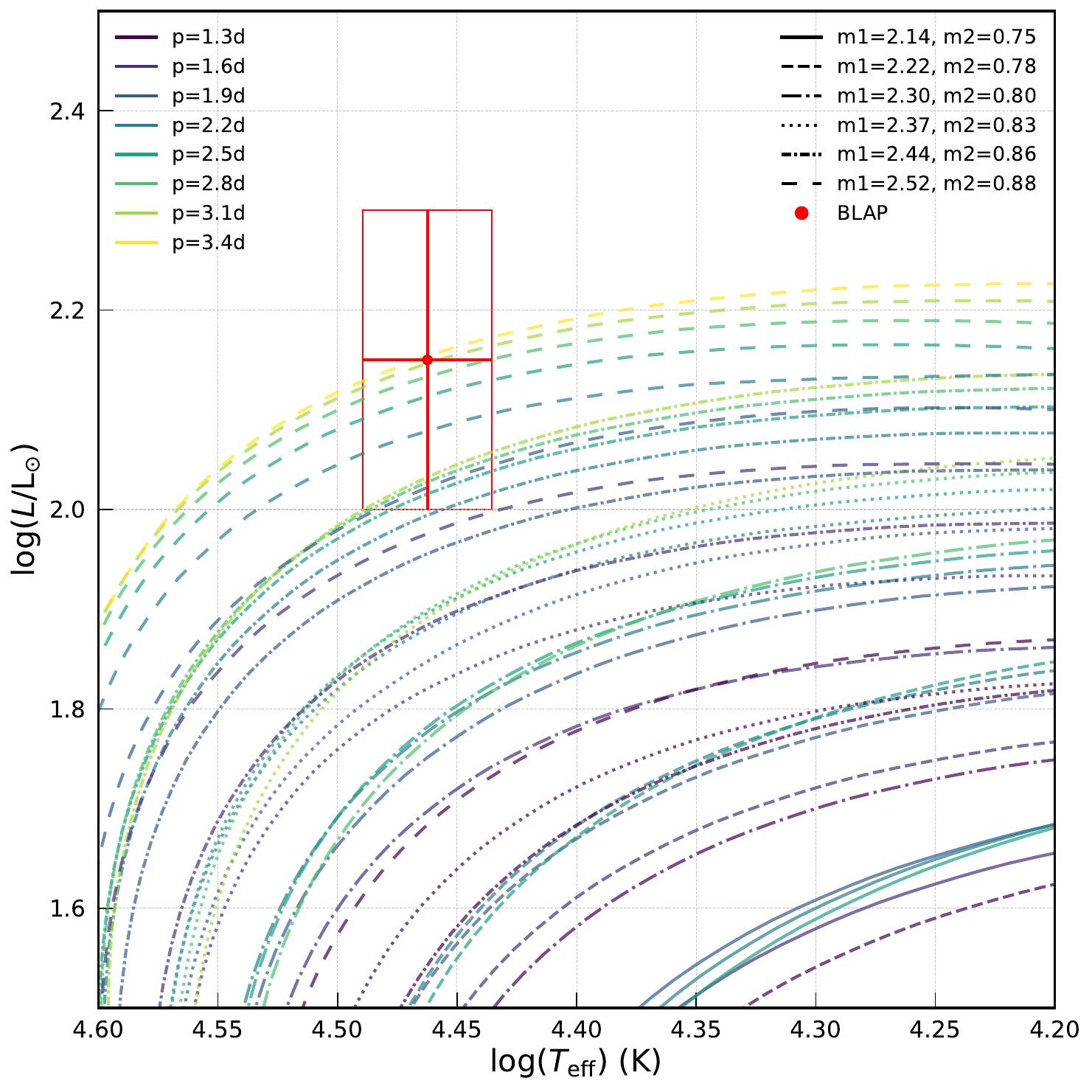}}
\end{center}
\caption{HR diagram for all RLOF channel models with a mass ratio of 0.35. Different line styles represent different binary masses, and different colors represent different initial orbital periods.}
\label{fig4}
\end{figure}

We simulated a binary system with a primary star of $\rm 2.14 \, M_{\sun}$ and a secondary star of $\rm 0.75 \, M_{\sun}$. The system has an orbital period of 1 day. In our simulations, we achieved stable mass transfer with a rate of approximately $10^{-6} \,\rm M_{\sun} \,yr^{-1}$ until the primary star’s mass decreased to $\rm 0.73 \, M_{\sun}$ and the secondary star’s mass increased to $\rm 2.0 \, M_{\sun}$. At that stage, the secondary star had exhausted its core hydrogen. However, observations of BLAP indicate that the companion is a MS star. This result shows that our model does not satisfy the observational requirements. Even when we increased the masses of both stars (up to their maximum plausible values), the secondary star still evolved off the MS, leading to a similar outcome.

The situation differs when the orbital period increases. Figure \ref{fig3} shows the evolution of systems with a primary star of $\rm 2.14 \, M_{\sun}$ and a secondary star of $\rm 0.75 \,M_{\sun}$. For initial orbital periods of $1.3$ to $2.5$ days, RLOF begins once the primary star exhausts its core hydrogen and enters the Hertzsprung gap. During mass transfer, the secondary star grows to a mass of $\rm 2.61-2.63 \, M_{\sun}$ while it remains on the MS. For systems with initial orbital periods between $2.8$ and $4.0$ days, the mass transfer rate quickly reaches $\rm 10^{-3.8}\, M_{\sun} \,yr^{-1}$. This high rate indicates that the binary system likely enters a CE phase as soon as mass transfer begins. Table 3 summarizes the complete evolutionary parameters for binary systems with a mass ratio $q=0.35$ undergoing RLOF.

Figure \ref{fig4} shows all cases with $q = 0.35$ in which the secondary star remains on the MS after mass transfer. Only the system with the highest initial masses ($M_{1}=2.52 \, \rm M_{\sun}$ and $M_{2}=0.88 \, \rm M_{\sun}$) and initial orbital periods between 2.2 and 3.4 days shows the primary star’s evolutionary track crossing the BLAP region. However, even when the primary star crosses the BLAP region, calculations show that the pulsation period remains below 32.37 minutes, which is inconsistent with observations. These results indicate that a  $q = 0.35$ cannot explain the observed BLAP.

\begin{table*}
\label{table3}
\caption{Parameters for stable mass transfer models at $q = 0.35$. $P_{\rm i}$ represents initial orbital periods enabling stable mass transfer, $M_{\rm 1,f}$ is the final mass of $M_{1}$ after mass transfer, and $P_{\rm f}$ is the orbital period at the end of mass transfer.}
\centering
\begin{tabular}{c c c c c}
\hline\hline
$M_{\rm 1,i}(\rm M_{\sun})$ & $M_{\rm 2,i}(\rm M_{\sun})$ & $P_{\rm i}(\rm days)$ & $M_{\rm 1,f}(\rm M_{\sun})$ & $P_{\rm f}(\rm day)$ \\
\hline
2.14 & 0.75 & 1.3-2.5 & 0.252-0.277 & 17.72-27.11 \\
2.22 & 0.78 & 1.3-2.5 & 0.259-0.285 & 18.42-27.97 \\
2.30 & 0.80 & 1.3-2.8 & 0.266-0.297 & 18.54-30.31 \\
2.37 & 0.83 & 1.3-3.1 & 0.272-0.307 & 19.26-33.62 \\
2.44 & 0.86 & 1.3-3.1 & 0.269-0.316 & 19.57-34.18 \\
2.52 & 0.88 & 1.3-3.4 & 0.274-0.327 & 19.69-36.31 \\
\hline
\end{tabular}
\end{table*}

\subsection{Cases for $q = 0.45$}
The binary evolution at this mass ratio is similar to the evolution observed for $q = 0.35$. Assuming that the initial orbital period is $P_{\rm i}(\rm days) = 1.0$ for all masses, the primary star begins mass transfer during its MS phase. Accretion by the secondary star accelerates its evolution. Consequently, by the time the secondary exhausts its core hydrogen, the primary star still retains approximately $0.03\rm \, M_{\sun}$ of hydrogen. This finding is inconsistent with the observed evolutionary states of BLAP.

Mass transfer via RLOF can occur at longer initial orbital periods. We consider a representative system having initial masses $M_{\rm 1,i}=2.00 \rm \, M_{\sun}$ and $M_{\rm 2,i}=0.90 \rm \, M_{\sun}$. For initial orbital periods $P_{\rm i}$ between 1.3 and 2.2 days, mass transfer is initiated when the primary star leaves the MS and enters the Hertzsprung gap. In contrast, the secondary remains on the MS. As a result, mass transfer reduces the primary to a pre–WD with a mass in the range $0.265-0.282 \rm \, M_{\sun}$, while the secondary’s mass increases to between $2.618$ and $2.635 \rm \, M_{\sun}$; it also remains on the MS. However, the primary’s later evolutionary track on the HR diagram lies below the typical BLAP region.

For systems with initial orbital periods between 2.5 and 4.0 days, mass transfer begins during the Hertzsprung gap phase. However, the mass transfer rate quickly reaches approximately $\rm 10^{-3.76} \, M_{\sun} \, yr^{-1}$, indicating the onset of a CE phase. Increasing the total system mass produces results similar to those in Figure \ref{fig4}; even the highest-mass configurations do not yield pulsation periods that match the observed BLAP characteristics. Table 4 summarizes all instances in which RLOF occurs and the secondary star remains on the MS following mass transfer.

\begin{table*}
\label{table4}
\caption{Parameters for stable mass transfer at $q = 0.45$. $P_{\rm i}$ represents initial orbital periods enabling stable mass transfer, $M_{\rm 1,f}$ is the final mass of $M_{1}$ after mass transfer, and $P_{\rm f}$ is the orbital period at the end of mass transfer.}
\centering
\begin{tabular}{l c c c c}
\hline\hline
$M_{\rm 1,i}(\rm M_{\sun})$ & $M_{\rm 2,i}(\rm M_{\sun})$ & $P_{\rm i}(\rm days)$ & $M_{\rm 1,f}(\rm M_{\sun})$ & $P_{\rm f}(\rm day)$ \\
\hline
2.00 & 0.90 & 1.3-2.2 & 0.265-0.282 & 21.24-31.30 \\ 
2.07 & 0.93 & 1.3-2.2 & 0.262-0.283 & 24.18-33.97 \\
2.14 & 0.96 & 1.3-2.5 & 0.263-0.288 & 26.20-40.56 \\
2.21 & 0.99 & 1.3-2.8 & 0.264-0.294 & 28.35-46.82 \\
2.28 & 1.02 & 1.3-2.8 & 0.262-0.299 & 27.52-48.27 \\
2.35 & 1.05 & 1.3-3.1 & 0.265-0.309 & 28.76-53.21 \\
\hline
\end{tabular}
\end{table*}

\subsection{Cases for $q = 0.30$}
\subsubsection{From ZAMS to BLAP}

\begin{table*}
\label{table5}
\caption{Parameters for stable mass transfer at $q = 0.30$. $P_{\rm i}$ represents initial orbital periods enabling stable mass transfer, $M_{\rm 1,f}$ is the final mass of $M_{1}$ after mass transfer, and $P_{\rm f}$ is the orbital period at the end of mass transfer.}
\centering
\begin{tabular}{l c c c c}
\hline\hline
$M_{\rm 1,i}(\rm M_{\sun})$ & $M_{\rm 2,i}(\rm M_{\sun})$ & $P_{\rm i}(\rm days)$ & $M_{\rm 1,f}(\rm M_{\sun})$ & $P_{\rm f}(\rm day)$ \\
\hline
2.23 & 0.67 & 1.3-2.5 & 0.259-0.284 & 12.28-20.38 \\ 
2.31 & 0.69 & 1.3-2.5 & 0.266-0.293 & 13.48-20.13 \\
2.38 & 0.72 & 1.3-2.8 & 0.272-0.304 & 14.14-22.91 \\
2.46 & 0.74 & 1.3-3.1 & 0.278-0.316 & 14.33-24.80 \\
2.54 & 0.76 & 1.6-3.4 & 0.305-0.328 & 14.95-26.45 \\
2.62 & 0.78 & 1.6-3.7 & 0.314-0.339 & 14.89-28.25 \\
\hline
\end{tabular}
\end{table*}

For models with an initial orbital period of 1 day, mass transfer commences before the primary star completes core hydrogen burning, and the orbital period decreases during this process. At the orbital period minimum, the accretion rate abruptly increases to approximately $\rm 10^{-4} \, M_{\sun} \, yr^{-1}$, consistent with the case of $q = 0.25$ shown in Figure \ref{fig2}, suggesting the onset of a CE phase. For the two binary model sets with the highest initial masses (the last two in Table 5) and an initial orbital period of 1.3 days, the evolutionary path differs from that of models with $q = 0.35$ and $q = 0.45$, as RLOF does not occur. In these cases, mass transfer begins before the primary star completes core hydrogen burning, leaving between 0.04 and 0.06 of its core hydrogen unburnt. At the orbital period minimum, the accretion rate abruptly increases to approximately $\rm 10^{-3.9} \, M_{\sun} \, yr^{-1}$, potentially signaling the onset of a CE phase.

We focus on models with orbital periods longer than 1 day. As an illustrative example, consider the binary system with the lowest total mass, where $M_{1}=\rm 2.23 \, M_{\sun}$ and $M_{2}=\rm 0.67 \, M_{\sun}$. For an initial orbital period between 1.3 and 2.5 days, the stars complete core hydrogen burning and then enter the Hertzsprung gap phase. Subsequently, they fill their Roche lobes, initiating mass transfer via RLOF. Initially, the orbital period decreases until it reaches a minimum, after which the primary and secondary masses reverse. After this reversal, the orbital period increases, and the radius of the primary star shrinks significantly so that it no longer fills its Roche lobe. Consequently, mass transfer ceases. After mass transfer, $M_{1}$ is reduced to $\rm 0.259$–$\rm 0.284 \, M_{\sun}$, $M_{2}$ increases to $\rm 2.616$–$\rm 2.641 \, M_{\sun}$, and the orbital period expands to 12.28–20.38 days. The post-mass transfer parameters for more massive binaries are listed in Table 5. Models with orbital periods outside this range likely entered a CE phase because the initial mass transfer rate was excessively high. This behavior is similar to that shown in Figure \ref{fig3}.

\begin{figure}[ht!]
\begin{center}
\resizebox{\hsize}{!}{\includegraphics{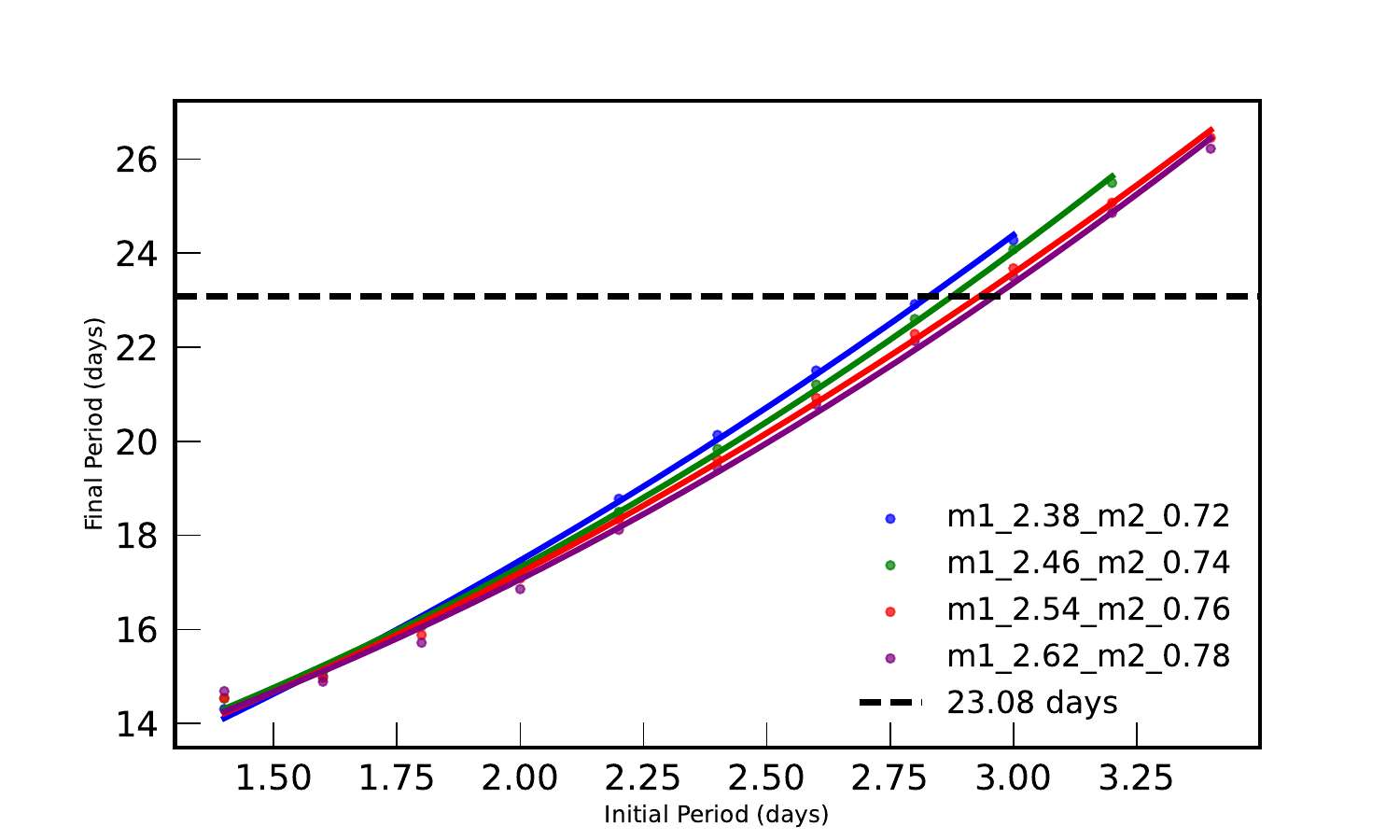}}    
\end{center}
\caption{Relationship between final and initial orbital periods for binary systems with mass ratio $q = 0.30$. Polynomial fits are applied to the four datasets. The horizontal dashed black line represents a final orbital period of 23.08 days.}
\label{fig5}
\end{figure}

\begin{figure*}[ht!]
\resizebox{\hsize}{!}{\includegraphics{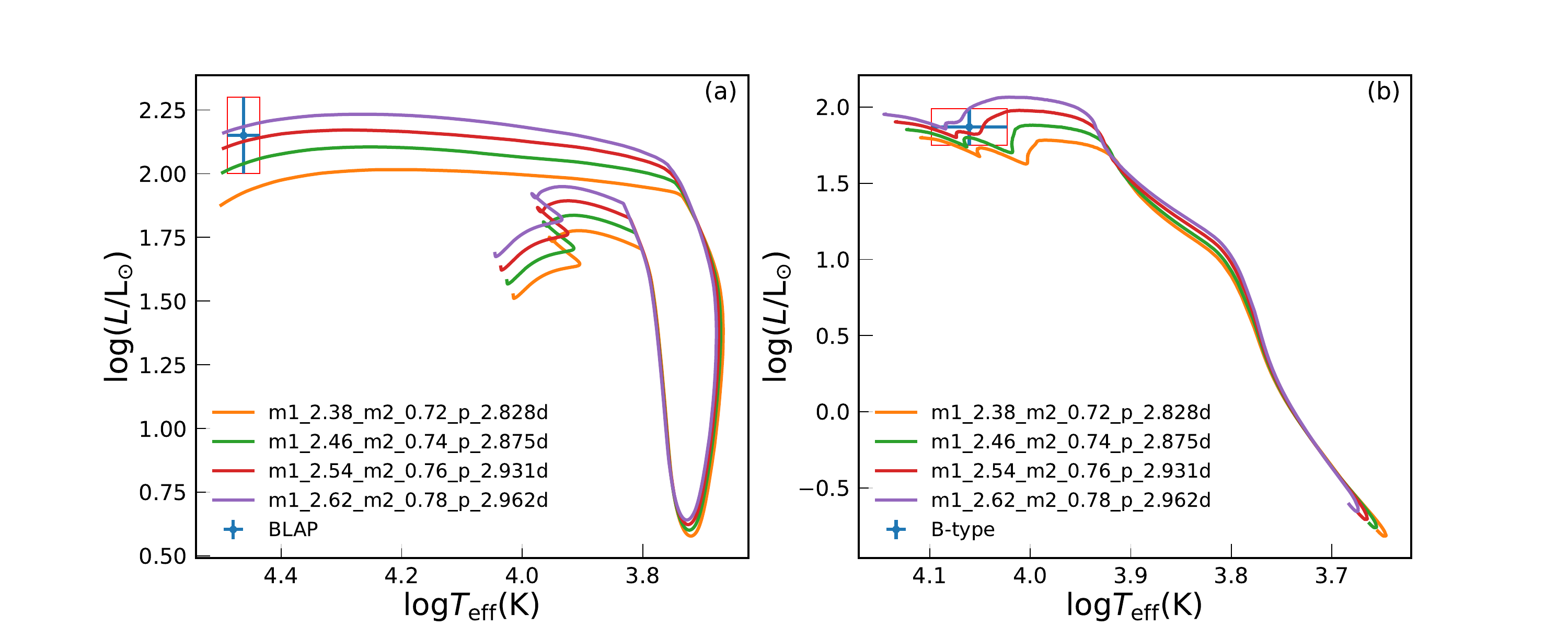}}
\caption{Evolutionary tracks on the HR diagram for binary systems with optimal initial orbital periods derived from polynomial fits. Panel (a) shows the evolutionary paths of the primary stars from ZAMS through the BLAP phase. Panel (b) displays the corresponding evolutionary tracks of the secondary stars.}
\label{fig6}
\end{figure*}

We identified a set of RLOF models; however, they do not yet accurately reproduce the observed final orbital and pulsation periods. Therefore, we refined the parameters of our RLOF models. The observed orbital period of the BLAP is 23.08 days. To identify the best-fitting models, we plotted the initial versus final orbital periods for the last four model sets (Table 5), as shown in Figure \ref{fig5}, and applied a polynomial fit to determine the best-fit orbital period. We identified four models whose final orbital periods are approximately 23.08 days. Subsequently, we plotted these binary systems on the HR diagram (Figure \ref{fig6}). We found that models with $M_{1}$ masses of $\rm 2.46 \, M_{\sun}$, $\rm 2.54 \, M_{\sun}$, and $\rm 2.62 \, M_{\sun}$ all pass through the BLAP region. However, the effective temperatures of the MS stars are higher than the observed values. We will address this discrepancy in the next section.

\begin{figure*}[ht!]
\resizebox{\hsize}{!}{\includegraphics{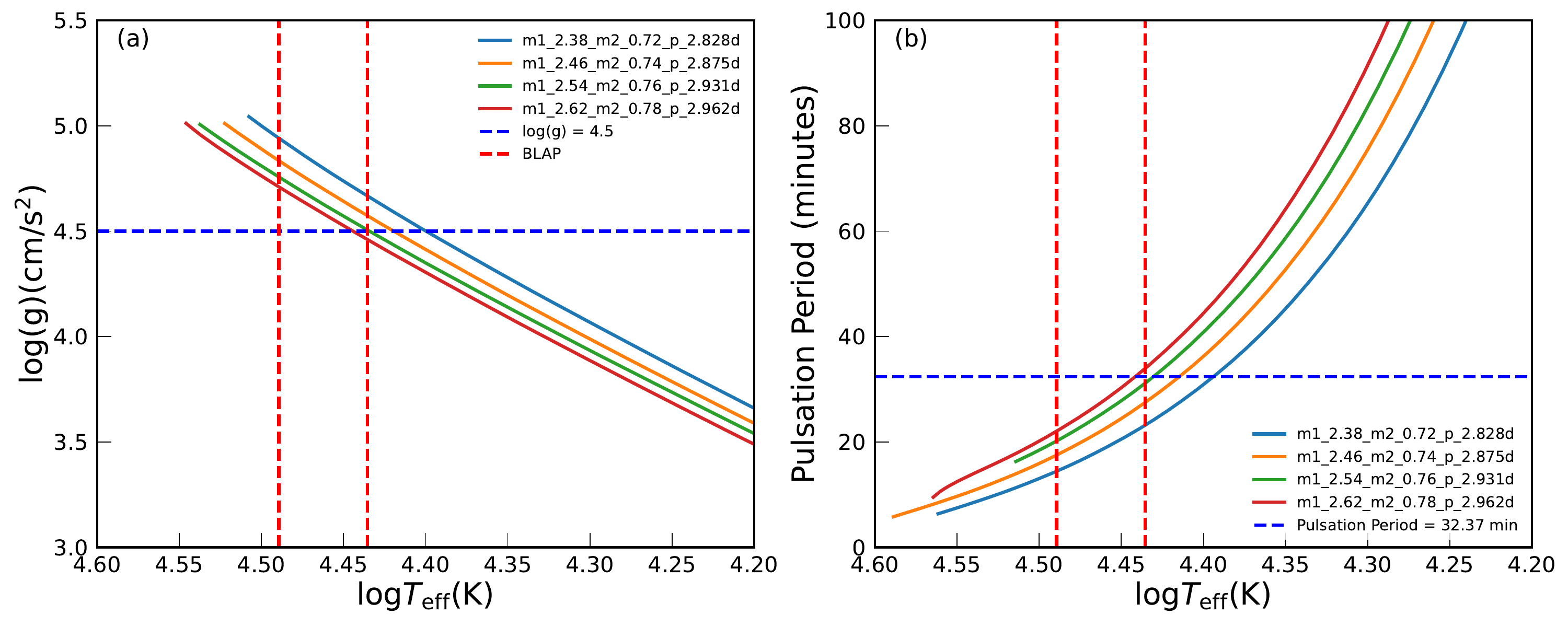}}
\caption{Diagnostic diagrams showing the evolution of stellar parameters for four binary models selected based on their final orbital period. Panel (a) presents the relationship between effective temperature and surface gravity, with horizontal and vertical dashed lines indicating the observational constraints for BLAP (red: ${\rm log}T_{\rm eff}$ range; dark blue: ${\rm log}(g/\rm cm \, s^{-2}) = 4.5$). Panel (b) shows the correlation between effective temperature and pulsation period, where the horizontal and vertical dashed lines mark the observed BLAP temperature range (red) and the target pulsation period of 32.27 minutes (dark blue), respectively.}
\label{fig7}
\end{figure*}

As shown in Figure \ref{fig7}, two models, with $M_{1} = \rm 2.54 \, M_{\sun}$ and $M_{1} = \rm 2.62 \, M_{\sun}$ respectively, fall within the ${\rm log}T_{\rm eff}$ range of BLAPs and satisfy the corresponding ${\rm log}(g)$ criteria. However, only the model with $M_{1} = \rm 2.62 \, M_{\sun}$ reaches a pulsation period of 32.37 minutes within the ${\rm log}T_{\rm eff}$ region characteristic of BLAPs. Next, we compare the relative rate of period change. The observed value is given by $r \equiv \dot{P}_{\rm P}/P_{\rm P} = (-11.5 \pm 0.6) \times 10^{-7} \,\rm yr^{-1}$. At a pulsation period of 32.37 minutes, the calculated value is $r \equiv \dot{P}_{\rm P}/P_{\rm P} = -1.9 \times 10^{-6} \,\rm yr^{-1}$. These two values differ by approximately one order of magnitude.

This model offers a compelling explanation for the observed properties of BLAP. First, the primary star evolves through the BLAP region on the HR diagram; during this process, the model predicts that the pulsation period reaches 32.27 minutes. Second, the primary star attains a surface gravity of ${\rm log}(g/\rm cm \, s^{-2}) = 4.5$ during its traverse of the HR diagram. Third, the model predicts a relative rate of period change of $\rm -1.9 \times 10^{-6} \, yr^{-1}$ at a pulsation period of 32.27 minutes.

\subsubsection{Consider the $\beta$ value in the mass transfer processes.}
Given that the  ${\rm log}T_{\rm eff}$ of the B-type MS star in the previous model calculations was approximately 1400K higher than the observed value (Figure \ref{fig6}), we hypothesized that the mass transfer process was non-conservative. This speculation is reasonable and allows us to define \begin{equation}\beta^{\star}=-\frac{\dot{M}_a}{\dot{M}_d}=\min\left(\frac{CM_a/\tau_{KH, a}}{M_d/\tau_{KH,d}},1\right), \end{equation} 
where $\tau_{KH}$ is the Kelvin-Helmholtz timescale, $\dot{M}_d$ is the mass-loss rate of the donor star, $\dot{M}_a$ is the mass-accretion rate onto the accretor, and C accounts for the increase in the maximum accretion rate due to the expansion of the accreting star \citep[e.g.][]{1972AcA....22...73P, 1977PASJ...29..249N, 2002MNRAS.329..897H, 2015ApJ...805...20S}. During Case A mass transfer, if the component stars have roughly comparable masses, then $\rm \beta^{\star} \approx 1$. During Case B/C mass transfer - when the donor star is more evolved – $\beta^{\star}$ is typically closer to 0. It is evident that the mass transfer in our model is Case B, and $\rm \beta^{\star}$ can be different from 1 (note that the definition of $\rm \beta^{\star}$ here is the opposite of that in MESA).

\begin{figure*}[ht!]
\resizebox{\hsize}{!}{\includegraphics{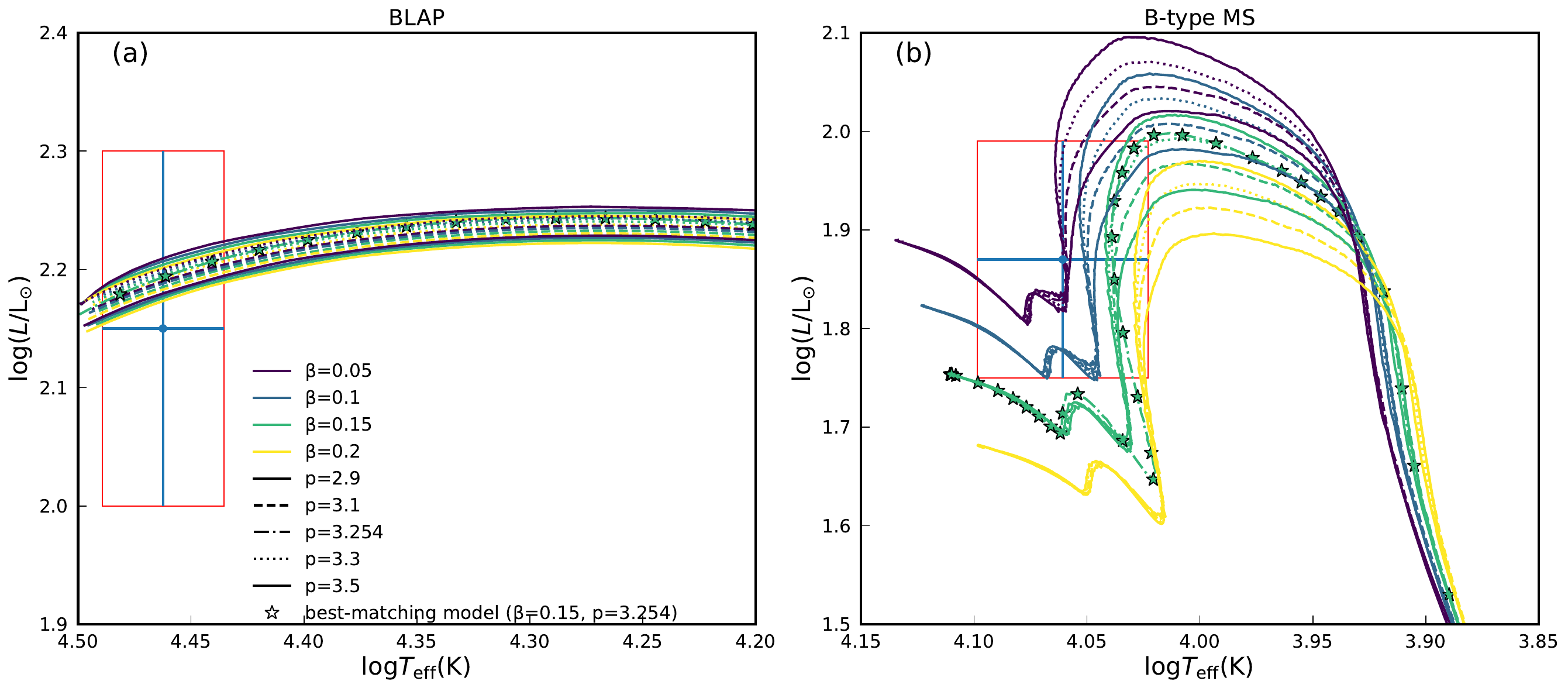}}
\caption{HR diagram showing the influence of $\beta$ on the evolutionary tracks of both binary components. Panel (a) displays the evolutionary paths of the primary stars for various $\beta$ values and initial orbital periods, with the pentagram marking the model that best reproduces the observed orbital period of 23.08 days. Panel (b) illustrates the corresponding evolutionary tracks of the secondary stars. Different colors represent different $\beta$ values, while different line styles indicate varying initial orbital periods.}
\label{fig8}
\end{figure*}

We performed calculations that incorporated the parameter $\beta$. We initialized the primary and secondary stars with masses of $M_{1} = 2.62 \, \rm M_{\sun}$ and $M_{2} = 0.78 \, \rm M_{\sun}$, respectively. The initial orbital period ranged from 2.9 to 3.5 days, and we adopted a time step of 0.2 days. We explored several values of $\beta$ (0.05, 0.10, 0.15, and 0.20) to investigate their influence on the system’s evolution. Adopting the method described in the $q = 0.30$ subsection for identifying the best-fitting final period, we determined the optimal model to have $P_{\rm i}(\rm days) = 3.254$. As shown in Figure \ref{fig8}, the evolutionary tracks indicate that the position of the B-type MS star on the HR diagram is closely related to the $\beta$ value. Increasing the value of $\beta$ leads to lower luminosity and effective temperature for the MS star. At $\beta=0.15$, the calculated effective temperature differs from the observed value by approximately 300K, and the luminosity falls within the observational error range. This non-conservative mass transfer scenario offers a more compelling explanation for the observed properties of the B-type MS star. Furthermore, we calculated that the binary system requires approximately $\rm 4.44 \times 10^{8} \, yr$ to evolve from the MS to the BLAP stage. Subsequently, it remains in the BLAP region for about $\rm 2.19 \times 10^{5} \, yr$.

\section{Disscusion}\label{diss}

\subsection{Post-BLAP Evolution}
\begin{figure*}[ht!]
\resizebox{\hsize}{!}{\includegraphics{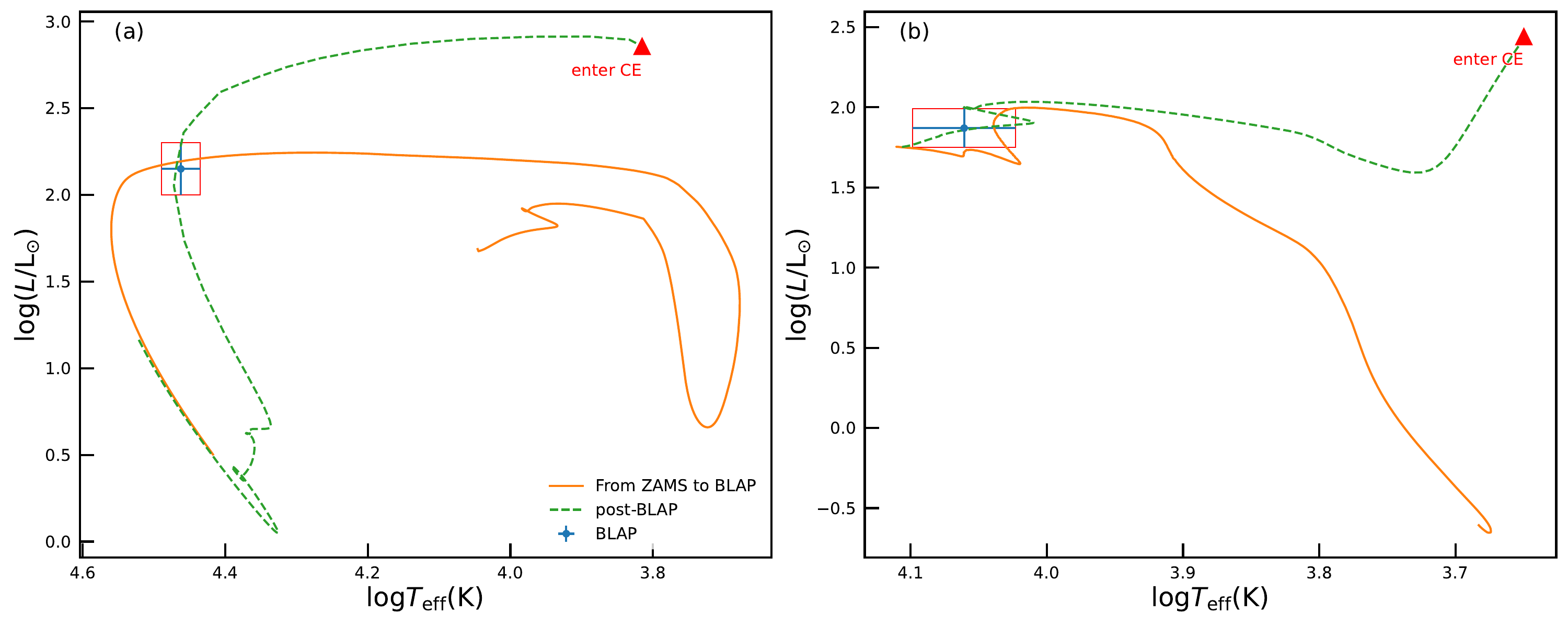}}
\caption{Complete evolutionary tracks for both components in the binary system from ZAMS through the BLAP phase and their subsequent evolution until the onset of CE. Panel (a) shows the evolutionary path of the primary star as it transitions from ZAMS through the BLAP phase (orange solid line) and its subsequent evolution toward the CE phase (green dashed line), with the red triangle marking the point of CE initiation. Panel (b) presents the corresponding evolution of the secondary star, illustrating its evolution until the system enters the CE phase, following the same line style and color conventions as Panel (a).}
\label{fig9}
\end{figure*}

After completing the BLAP phase for the model that best matches observations in Figure \ref{fig8}, we continued the evolutionary calculations for this binary system. Next, the B-type MS star becomes the mass donor, while the former BLAP serves as the accretor. As the donor evolves to the tip of the red giant branch, it fills its Roche lobe and transfers mass to its companion. At this stage, the mass ratio is as low as 0.12, making stable mass transfer unlikely ($\dot{M} \approx 10^{-3} \rm \, M_{\sun}\, yr^{-1}$). We therefore speculate that the system has now entered the CE phase. It takes approximately $\rm 5.5 \times 10^{8} \, yr$ after the BLAP phase to reach the CE phase. The HR diagram of the post-BLAP stage is shown in Figure \ref{fig9}.

The binding energy of the MS star was calculated to be $\rm -5.43 \times 10^{48} \, erg$. In an extreme scenario of CE ejection, the limiting separation of the two stars is given by $R_{\rm 1,core} + R_{2} = 0.294 \, \rm R_{\sun}$. In contrast, the separation before entering the CE phase is $49 \rm R_{\sun}$. As a result, the orbital energy decreases by $\rm 7.61\times 10^{47} \, erg$. Based on these energy values, the calculated $\rm \alpha_{CE}$ is approximately 7.14. The widely accepted range for $\rm \alpha_{CE}$ is approximately 0 to 1.5. This discrepancy suggests that the orbital energy is likely insufficient to unbind the common envelope, leading to the conclusion that the two stars will ultimately merge.

\subsection{Elemental Abundances on the Surfaces of MS}
In binary systems experiencing mass transfer, the accretion of material from a donor star onto its companion fundamentally alters the surface abundances of the accreting star. For example, helium enrichment directly indicates mass accretion from a helium-rich donor star. Likewise, enhanced nitrogen abundance relative to carbon and oxygen signals the accretion of CNO-processed material. These elemental signatures are crucial for understanding the history of mass transfer and tracing evolutionary pathways.

\begin{figure}[ht!]
\begin{center}
\resizebox{\hsize}{!}{\includegraphics{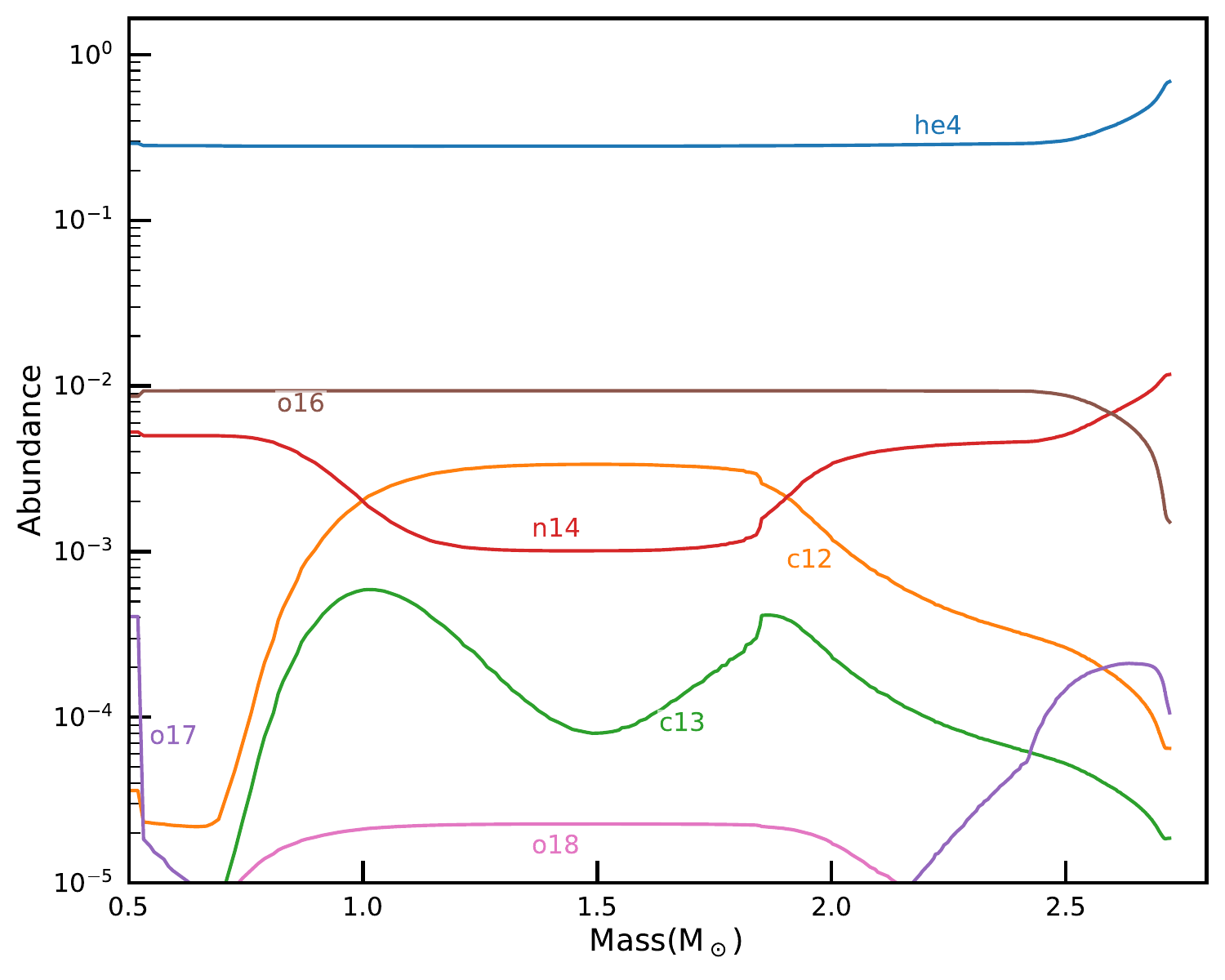}}
\end{center}
\caption{Distribution of elemental abundances in the MS star after mass transfer, with distinct enhancements in $^{4}\rm He$ and $^{14}\rm N$ abundances.}
\label{fig10}
\end{figure}

Our MESA evolutionary calculations provide quantitative predictions for the photospheric abundances of the B-type MS star. Figure \ref{fig10} illustrates that our simulations reveal significant enhancements in helium and nitrogen abundances at the surface of the mass-accreting MS star. Specifically, the helium mass fraction reaches $X_{\rm He} = 0.68$, while the nitrogen abundance is elevated to $X_{\rm N} = 0.01$. The derived surface composition exhibits characteristic signatures consistent with the accretion of CNO-processed material originating from an evolved donor. Furthermore, our analysis of the isotopic abundance ratios reveals: $^{13}C/^{12}C = 0.286886$, $^{14}N/^{15}N = 23873$, $^{17}O/^{16}O = 0.070507$, and $^{18}O/^{16}O = 0.000011$.

However, the lack of detailed spectroscopic analysis for HD 133729 presents a challenge for directly validating our theoretical models. Future high-resolution spectroscopic studies are crucial for accurately measuring the surface abundances of He and N in the B-type MS star. Confirmation of these abundance anomalies would strongly support the viability of this evolutionary pathway.

\subsection{Adiabatic calculation of GYRE}

\begin{figure}[ht!]
\begin{center}
\resizebox{\hsize}{!}{\includegraphics{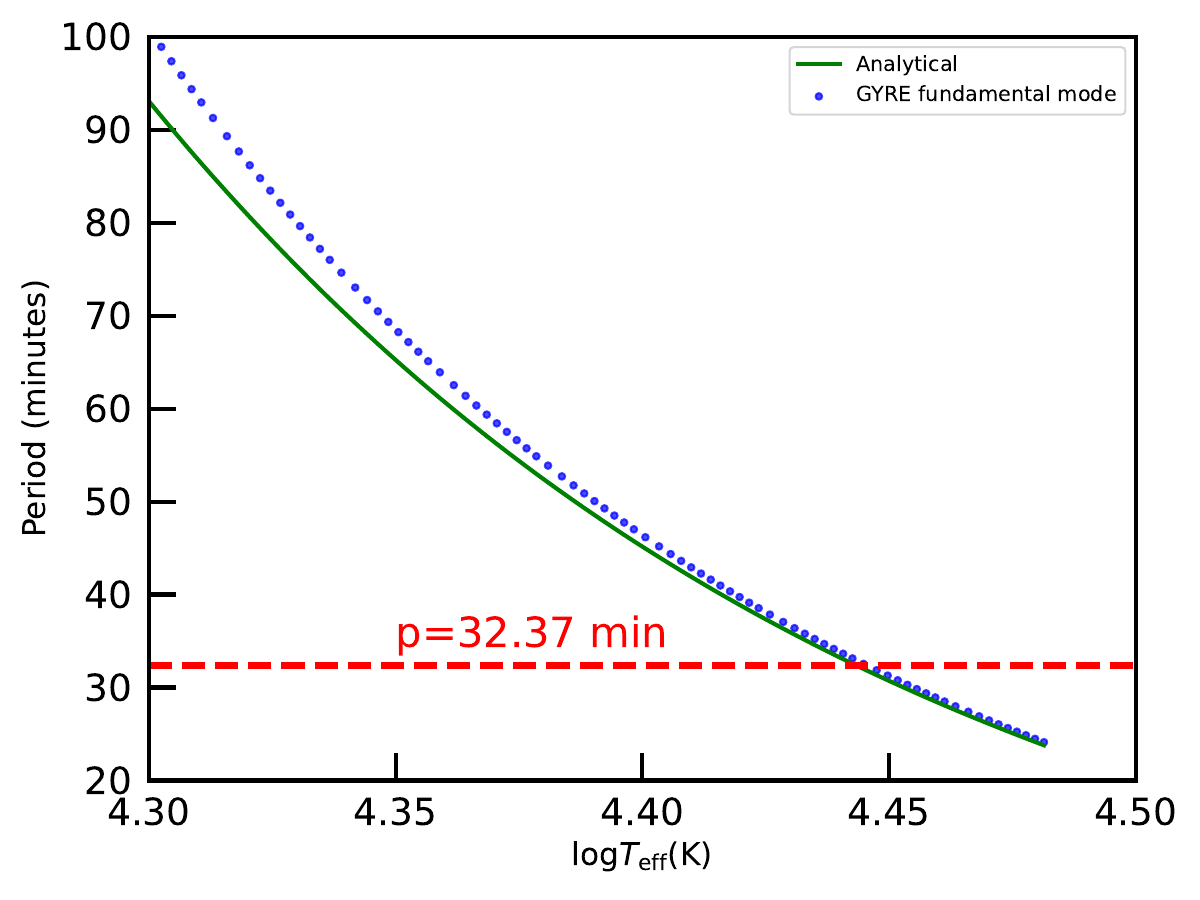}}
\end{center} 
\caption{The ${\rm log}T_{\rm eff}$ vs. pulsation period diagram for the best-match model shown in Figure \ref{fig8}. The green solid line represents the results calculated using Equation 1, while the blue dotted line shows the results of adiabatic pulsation calculations using GYRE.}
\label{fig11}
\end{figure}

\citet{2019ApJ...878L..35K} used formula (1) to calculate the pulsation period of high-gravity BLAPs. We aim to validate this formula for normal BLAPs. We calculated the fundamental mode pulsation periods of BLAP near 32.75 minutes using adiabatic mode computations with GYRE, as shown in Figure \ref{fig11}. Around 32.75 minutes, the formula provides an excellent match to the GYRE computations, further demonstrating the efficacy of this formula for a wide range of BLAPs.

\subsection{Pulsational instability uncertainties}
\citet{2020MNRAS.492..232B} investigated the pulsational stability of pre-WD models during their evolution. They performed adiabatic analyses to identify the eigenfrequencies of the stars. These adiabatic results were then used as initial guess values for nonadiabatic analyses to determine the stability of the identified modes. They found that the radial fundamental mode is unstable from an effective temperature of roughly 30,000K to at least 50,000K, and possibly even as high as 80,000K. In this temperature range, the periods of the unstable fundamental mode vary from around 100 seconds to as long as 2-3 hours.

\citet{2018MNRAS.477L..30R} also found unstable modes for three harmonic degrees. g modes with $l$ = 1 and $l$ = 2 are unstable for radial orders in the range $29 \le k \le  39$ and $54 \le k \le  67$, respectively. The radial fundamental mode is also unstable, suggesting that the modes observed in BLAPs with periods of 1200-1500 seconds could be radial modes. The observed periods in BLAPs can be explained by non-radial g modes with high radial order or by low-order radial modes in the case of the shortest periods.

\subsection{MESA-RSP}
The MESA-RSP (Radial Stellar Pulsations) software package, integrated into MESA, has become a powerful tool for studying stellar pulsations. This package can reliably simulate large-amplitude, self-excited, nonlinear pulsations in classical pulsating stars, making it particularly valuable for studying BLAPs. \citet{2024arXiv240816912J} pioneered the first comprehensive exploration of BLAP theoretical instability regions using MESA-RSP, constructing an extensive grid of models with masses ranging from $\rm 0.3-1.1 \, M_{\sun}$, luminosities spanning ${\rm log}(L/\rm L_{\sun}) = 1.5-3.5$, and effective temperatures between 20-35 kK. Their analysis revealed that for lower luminosities ($\rm \sim 200L_{\sun} $), as observed by \citet{2022A&A...663A..62P} and \citet{2024MNRAS.52710239B}, the fundamental mode periods of lower-mass models ($\rm 0.3-0.4 \, M_{\sun}$) show better agreement with observations and empirical relations, particularly the period-luminosity relation.

\subsection{Broader applications of the model}
Previous study presents the orbital period and companion mass distribution for stars in the BLAP phase (see Figure 9 of \citet{2021MNRAS.507..621B}). The orbital period exhibits a bimodal distribution, with a minor peak around 1.2 days and a more prominent peak at an orbital period of $\sim$40 days. Based on the orbital period of HD 133729 and our calculations, we propose that the second peak in the orbital period distribution likely arises from RLOF, while the first peak originates from the CE ejection. Our aim is not only to explain HD 133729 but also to provide a broad grid of calculated parameters. 

\section{Summary}\label{sum}
This study systematically explores the origin and evolution of binary BLAP systems, with a particular focus on HD 133729. We investigated a broad parameter space of mass ratios ($q$ = 0.25, 0.30, 0.35, 0.45) and initial orbital periods, revealing that the pre-WD RLOF channel is the most viable formation pathway for HD 133729. Our simulations show that while a conservative mass transfer model with $q = 0.30$ reproduces many of the observed properties (orbital period, pulsation period, and rate of period change), it overestimates the effective temperature of the B-type MS star.

To reconcile this discrepancy, we considered non-conservative mass transfer with an isotropic re-emission wind efficiency of $\beta = 0.15$, which yielded better agreement with the observed temperature. The best-fit model for HD 133729 has component masses of $M_{\rm 1,i} = 2.68 \, \rm M_{\sun}$ and $M_{\rm 2,i} = 0.78 \,\rm M_{\sun}$, and an initial orbital period of $P_{\rm i} = 3.254 \, \rm day$. Furthermore, we predict that the system will undergo a CE phase during the second mass transfer episode, ultimately leading to a merger.

Our results also highlight the importance of surface abundance evolution. After the first mass transfer phase, we expect notable helium and nitrogen enhancements in the B-type MS companion due to accretion of CNO-processed material. Such abundance signatures can be directly tested with high-resolution spectroscopy. Furthermore, we have found that non-conservative mass transfer plays a crucial role in the evolution of BLAPs. This provides a novel approach for constraining mass loss during mass transfer in binary systems, potentially offering insights beyond the scope of traditional asteroseismology.

\section{Acknowledgments}
We would like to thank Thomas Kupfer, Andrzej Pigulski, and Evan B. Bauer for their help. This study is supported by the National Natural Science Foundation of China (Nos 12288102, 12225304, 12090040/12090043, 12473032), the National Key R\&D Program of China (No. 2021YFA1600404), the Western Light Project of CAS (No. XBZG-ZDSYS-202117), the science research grant from the China Manned Space Project (No. CMS-CSST-2021-A12), the Yunnan Revitalization Talent support Program (Yunling Scholar Project), the Yunnan Fundamental Research Project (No 202201BC070003), and the International Centre of Supernovae, Yunnan Key Laboratory (No. 202302AN360001).

%---------------------------------------------
% References
\bibliographystyle{aa}
\bibliography{work2}

%---------------------------------------------
% Appendix

% End of the document
\end{document}